\documentclass[a4paper,11pt]{article}
\pdfoutput=1 

\usepackage{jheppub}
\usepackage[T1]{fontenc} 

\usepackage[dvipsnames]{xcolor}
\usepackage{tikz}
\usetikzlibrary{patterns}
\usetikzlibrary{snakes}

\newcommand{\phase}{\sigma}

\newcommand{\PhiSG}{\varPhi}

\newcommand{\bal}{\begin{equation}\begin{aligned}}
\newcommand{\eal}{\end{aligned}\end{equation}}

\newcommand{\ov}{\over}
\newcommand{\g}{\gamma}

\definecolor{grey}{rgb}{0.4,0.4,0.5}
\definecolor{darkgreen}{rgb}{0,0.5,0}
\definecolor{darkred}{rgb}{0.6,0.0,0}
\definecolor{lightbrown}{rgb}{1,0.9,0.8}
\definecolor{brown}{rgb}{0.6,0.3,0.3}
\definecolor{darkblue}{rgb}{0,0,0.5}
\definecolor{darkmagenta}{rgb}{0.5,0,0.5}

\newcommand{\ads}{${\rm  AdS}_5\times {\rm S}^5\ $}

\def\bQ{{\overline Q}}
\def\bY{{\overline Y}}
\def\bR{{\mathbb R}}

\newcommand{\la}{\label}
\def\a {\alpha}
\def\St{{\widetilde S}}
\def\Sb{{\bar S}}
\def\tSi{\widetilde \Sigma}
\def\Si{ \Sigma}
\def\bes{{\text{\tiny BES}}}
\def\tp{{\widetilde p}}
\def\tE{\widetilde\E}
\def\br {\bar\rho}
\def\tK{{\widetilde K}}
\def\hstar{\,\hat{\star}\,}
\def\cstar{\,\check{\star}\,}
\newcommand{\E}{\mathcal E}
\def\da {{\dot\alpha}}

\title{\boldmath Mirror Thermodynamic Bethe Ansatz for~AdS3/CFT2}

\author[a,1]{Sergey Frolov,%
\note{Correspondent fellow at Steklov Mathematical Institute, Moscow.}}
\author[b,c,2]{Alessandro Sfondrini%
\note{IBM Einstein Fellow.}}

\affiliation[a]{School of Mathematics and Hamilton Mathematics Institute,\\
Trinity College, Dublin 2, Ireland}
\affiliation[b]{Institute for Advanced Study,\\
Einstein Drive, Princeton, New Jersey, 08540 USA}
\affiliation[c]{Dipartimento di Fisica e Astronomia, Universit\`a degli Studi di Padova,\\
\& Istituto Nazionale di Fisica Nucleare, Sezione di Padova,\\
via Marzolo 8, 35131 Padova, Italy}

\emailAdd{frolovs@maths.tcd.ie}
\emailAdd{alessandro@ias.edu}

\abstract{We consider superstrings on the pure-Ramond-Ramond $AdS_3\times S^3\times  T^4$ background. Using the recently-proposed dressing factors for the worldsheet S matrix, we formulate the string hypothesis for the mirror Bethe-Yang equations, and use it to derive the canonical mirror thermodynamic Bethe ansatz (TBA) equations of the model.
For the first time, these equations account for all massive and massless modes of the model, and do not resort to any limit or special kinematics. We also discuss the simplified mirror TBA equations and the Y-system of the model. 
}

\begin{document} 
\maketitle
\flushbottom

\section{Introduction} 
\label{sec:introduction}

The ability to exactly and efficiently compute non-protected observables is crucial to understand and test a model. This is all the more important in the context of understanding AdS/CFT correspondence~\cite{Maldacena:1997re,Witten:1998qj,Gubser:1998bc}, which is a weak-strong duality. A remarkable achievement in this context was the computation of the spectrum of short-string states for freely-propagating $AdS_5\times S^5$ superstrings by integrability (see~\cite{Arutyunov:2009ga,Beisert:2010jr} for reviews), which is equivalent to the solution of the spectral problem for the $\mathcal{N}=4$ supersymmetric Yang-Mills theory in the planar limit~\cite{'tHooft:1973alw}.
The route to solving the spectral problem is clearer on the stringy side of the duality --- though historically many important insights were first obtained in gauge theory~\cite{Minahan:2002ve}. Considering the lightcone gauge-fixed Green-Schwarz action of the model as a two-dimensional integrable quantum field theory (IQFT), the AdS/CFT spectral problem can be reframed as the problem of determining the finite-volume spectrum of an IQFT. This is by no means an easy problem (especially as this particular IQFT is not Poincar\'e-invariant) but it is amenable to thermodynamic Bethe ansatz (TBA) techniques~\cite{Yang:1968rm,Zamolodchikov:1989cf,Dorey:1996re}. The first step is to identify an S matrix for the infinite-volume model~\cite{Beisert:2005tm}; a particularly challenging ingredient is the overall prefactor of the matrix, the so-called ``dressing factor''~\cite{Beisert:2006ez}, which is only partially constrained by crossing invariance~\cite{Janik:2006dc}.
Then, it is necessary to introduce a mirror model~\cite{Arutyunov:2007tc}, whereby the worldsheet time and space directions are swapped. Using this model, it is possible to describe the finite-volume ground-state energy of the original model as the (finite-temperature) free energy of the mirror model~\cite{Arutyunov:2009zu,Gromov:2009tv,Bombardelli:2009ns,Arutyunov:2009ur, Arutyunov:2009ux}. Excited states are obtained by analytic continuation~\cite{Dorey:1996re}. These mirror TBA equations can be substantially simplified and refined~\cite{ Arutyunov:2009ux,Arutyunov:2009ax,Cavaglia:2010nm} to  obtain eventually a more compact set of equations, which goes under the name of quantum spectral curve~\cite{Gromov:2013pga}.

We now want to follow this template for the case of the AdS3/CFT2 correspondence. Let us first highlight some fundamental differences between $AdS_5\times S^5$ and the case at hand. First of all, the maximally supersymmetric backgrounds depend on more parameters in the case of interest. For the purpose of computing the spectrum, there are \textit{two} (rather than \textit{one}) important parameters: the string tension, and the ratio of Ramond-Ramond (RR) \textit{versus} Neveu-Schwarz-Neveu-Schwarz (NSNS) background fluxes~\cite{OhlssonSax:2018hgc}. For any value of the fluxes, the backgrounds are known to be classically~\cite{Cagnazzo:2012se} and are argued to be quantum~\cite{Lloyd:2014bsa} integrable. Nonetheless, the models are quite different as we tune the ratio of background fluxes, and it is worth briefly discussing some of their features (see also~\cite{Sfondrini:2014via} for a review of early progress in AdS3/CFT2 integrability).

\paragraph{Pure-NSNS backgrounds.}
These are the simplest backgrounds as they are described by a Wess-Zumino-Witten model on the worldsheet~\cite{Maldacena:2000hw}. The spectrum involves both short strings and long strings, and it can be written down in a closed form. The short-string spectrum is highly degenerate, similarly to the spectrum of flat-space strings. Despite this apparent simplicity, only very recently some progress has been made in identifying the CFT2 duals~\cite{Giribet:2018ada,Eberhardt:2018ouy,Eberhardt:2021vsx}.
The spectrum can also be reproduced by integrability: the short-string spectrum follows from the mirror-TBA approach~\cite{Baggio:2018gct, Dei:2018mfl}, while the long-string sector needs an ad-hoc discussion~\cite{Sfondrini:2020ovj}.

\paragraph{Mixed-flux backgrounds.} 
Starting from the WZW description, it is possible to continuously deform the model introducing a coupling to RR background fluxes, see \textit{e.g.}~\cite{OhlssonSax:2018hgc}.
This immediately removes the long-string continuum from the spectrum~\cite{Eberhardt:2018vho} and lifts the degeneracies of short-string spectrum. It has been attempted~\cite{Eberhardt:2018exh} to use CFT techniques~\cite{Berkovits:1999im} to study this setup, but it does not seem easy to obtain the finite-coupling spectrum in that way. Integrability here may prove more efficient. An integrable S matrix for mixed-flux backgrounds is known~\cite{Hoare:2013pma,Lloyd:2014bsa}, though the dressing factors are not known and are probably quite intricate~\cite{Babichenko:2014yaa}. Unsurprisingly, this generic configuration of fluxes appears to be more challenging than either limit. Still, recent advances~\cite{Frolov:2021fmj, upcoming:mixed} suggest that it may be possible to construct the dressing factors and complete the integrability program for these backgrounds.

\paragraph{Pure-RR backgrounds.} 
This case is the one that resembles $AdS_5\times S^5$ most closely, and it is the one which we will consider in this paper. For a pure-RR background, the dispersion relation for a worldsheet excitation is
\begin{equation}
    E(p)=\sqrt{m^2 + 4h^2 \sin^2\big(\frac{p}{2}\big)}\,,
\end{equation}
and it is almost identical to $AdS_5\times S^5$. However, in that case, $m=1$, while here $m =\pm 1$ or~$m=0$~\cite{Borsato:2014exa}. In other words, here we have three distinct types of fundamental excitations, one of which is massless. This makes it hard to perform perturbative computations, as these massless modes yield IR divergences as well as a large contribution to finite-volume corrections~\cite{Abbott:2015pps}.
Another important difference is that the symmetry algebra of $AdS_5\times S^5$, which governs the form of its mirror TBA and quantum spectral curve, is given by the simple superalgebra $psu(2,2|4)$; by contrast, for $AdS_3\times S^3\times T^4$ we have instead $psu(1,1|2)\oplus psu(1,1|2)$. This makes it harder to guess some properties of the mirror TBA from symmetry considerations, see \textit{e.g.}~\cite{Gromov:2008gj}.
Nonetheless, the matrix part of the S~matrix for this background is known since 2014~\cite{Borsato:2012ud,Borsato:2013qpa,Borsato:2014hja}, and the dressing factors have been proposed around the same time~\cite{Borsato:2013hoa,Borsato:2016xns}. However, some of the analytic properties of that proposal were not entirely transparent, which may have hampered the derivation of the mirror TBA. Recently, we found another solution of the crossing equations~\cite{Frolov:2021fmj} which appears to have much nicer analytic properties and provides an excellent starting point for the derivation of the mirror~TBA.

The aim of this article is to derive the mirror TBA equations for the pure-RR $AdS_3\times S^3\times T^4$ background. Some attempts in this directions were already made in~\cite{Bombardelli:2018jkj,Fontanella:2019ury,Cavaglia:2021eqr,Ekhammar:2021pys}. In~\cite{Bombardelli:2018jkj}, a set of TBA equations was derived for a system obtained by first truncating the spectrum to the massless modes and then taking a relativistic limit. In~\cite{Fontanella:2019ury}, the authors considered again a truncation to massless modes, without taking any additional limit; however, they studied the finite-temperature, rather than finite-volume, theory. It is not obvious whether these constructions are related to the full spectrum and how.  One important issue is that the (mirror) TBA requires considering \textit{all} of the (mirror) model's expectations in the thermodynamic limit, and imposing thermodynamic equilibrium. As such, neglecting some types of particles may lead to drastically different results --- in other words, there are no closed subsectors in the finite-volume theory. More recently, the authors of~\cite{Cavaglia:2021eqr,Ekhammar:2021pys} have independently proposed the quantum spectral curve for a model with $psu(1,1|2)\oplus psu(1,1|2)$ symmetry. Their approach bypassed the derivation of the mirror TBA altogether, and instead relied on a symmetry bootstrap. It seems that the resulting equations capture some features of the $AdS_3\times S^3\times T^4$ Bethe-Yang equations; however, it is not clear if they account for the massless modes and how.%
\footnote{%
Additionally, the authors of~\cite{Cavaglia:2021eqr} claim that their QCS construction reproduces the dressing factors of~\cite{Borsato:2013hoa,Borsato:2016xns}, which are different from ours and, as explained in~\cite{Frolov:2021fmj}, are incompatible with parity invariance. It would be interesting to see how~\cite{Cavaglia:2021eqr} may be amended.
} 
Ultimately, the validity of the proposals~\cite{Cavaglia:2021eqr,Ekhammar:2021pys} will boil down to whether the quantum-spectral curve construction is uniquely constrained by symmetries, even for models whose symmetries are not given by a simple Lie (super)algebra, which is an outstanding question.

In light of the above discussion, we believe that the systematic derivation of the mirror TBA --- without any truncation or limit --- is necessary to understand the non-perturbative spectrum of $AdS_3\times S^3\times T^4$, and therefore we present it in this paper. This article is structured as follows. 
Firstly, working with the mirror model, we identify which fundamental and composite excitations contribute to the thermodynamic limit of the Bethe-Yang equations (large mirror volume, and large number of excitations). This discussion goes under the name of string hypothesis, and we present it in section~\ref{sec:stringhypothesis}. In the same section, we also reformulate the Bethe-Yang equations for such composite excitations by the fusion procedure. Using such equations, in section~\ref{sec:mirrortba} we derive the canonical mirror TBA equations and simplify them. We present our conclusions in section~\ref{sec:conclusions}. We have collected several technical results in the appendices: in appendix~\ref{app:definitions} we fix our parametrisations of the various particles appearing in the mirror model, in appendix~\ref{app:Smatrices} we defined the S matrices appearing in the Bethe-Yang equations (and the resulting integration kernels), and in appendix~\ref{app:factors} we collect the formulae relative to the dressing factors.

\section{String hypothesis} 
\label{sec:stringhypothesis}

The first step in the derivation of the mirror thermodynamic Bethe ansatz is studying the mirror model to determine which configurations of particles  appear in the thermodynamic limit (for large number of excitations and large mirror volume~$R$). These typically include real particles and their bound states, as well as complex configurations of particles and auxiliary roots. Collectively, all these configurations take the name of Bethe strings. The simplest setup where this sort of complex strings appear is perhaps the Heisenberg spin chain, see \textit{e.g.}~\cite{Faddeev:1996iy}. Even in such a simple setup, the detailed discussion of the thermodynamic limit is quite subtle, as there are non-string configurations that appear there \cite{Woynarovich:1981ca,Woynarovich:1982,Babelon:1982mc}. However, it can be shown that the model's free energy is correctly described by taking only string configurations into account \cite{Tsvelick:1983}. In most models, where the string configurations are usually quite a bit more intricate than for the Heisenberg chain, we formulate the hypothesis that all solutions relevant in the thermodynamic limit are Bethe strings. Under this hypothesis, it is not hard to identify the various string-like solutions to the mirror Bethe-Yang equations at large mirror volume~$R$. For $AdS_5\times S^5$ this was done in~\cite{Arutyunov:2009zu} (see also~\cite{vanTongeren:2013gva} for a more pedagogical presentation). Below we will follow and adapt that approach to the case of $AdS_3\times S^3\times T^4$.

\subsection{Mirror Bethe-Yang equations}
\label{sec:smirrorba}

Much like in $AdS_5\times S^5$~\cite{Arutyunov:2007tc}, we expect the mirror model to have bound states~\cite{Borsato:2013hoa,Frolov:2021fmj}. Moreover, these bound states are particularly simple when the Bethe-Yang equations are expressed in the ``$sl(2)$ grading''. In $AdS_5\times S^5$, where there is only one type of momentum-carrying modes, it is possible to express all Bethe-Yang equations in our grading of choice. For $AdS_3\times S^3\times T^4$, we can put \textit{e.g.}\ the ``left'' momentum-carrying modes%
\footnote{
Here the momentum carrying modes are distinguished by their eigenvalue under an $u(1)$ charge. For $M=+1$ we talk of left modes, for $M=-1$ we talk of right modes, and for $M=0$ we talk of massless modes.
} 
in our grading of choice. However, having done so, the grading of the remaining equations is fixed~\cite{Borsato:2012ss,Borsato:2013qpa,Borsato:2016xns}. While the choice of grading is eventually irrelevant, we find it convenient to write the equations for the ``left'' particles in the $sl(2)$ grading. As a consequence, the ``right'' particles will be in the $su(2)$ grading, and the massless particles will be in a fermionic grading.%
\footnote{It would be interesting to work out the eigenvalues of the transfer matrix of the mirror (as well as string) model in the appropriate grading~\cite{Seibold:2022mgg}.}

\paragraph{Fundamental excitations.}
Let us start by reviewing the different types of excitations that appear in the mirror Bethe-Yang equations for fundamental particles. We have $N_1$ ``left'' momentum carrying modes, and $N_{\bar{1}}$ ``right'' momentum carrying modes. The massless modes are in a doublet of an external $su(2)$, sometimes called $su(2)_{\circ}$~\cite{Borsato:2014exa}, so we introduce a label $\dot{\alpha}=1,2$ and the excitation numbers $N_{0}^{(\dot{\alpha})}$. Hence the total number of massless excitations is $N_0=N_0^{(1)}+N_0^{(2)}$. In principle, all momentum carrying modes may be described in terms of their mirror momentum $\tilde{p}_k$, but it is convenient to introduce a rapidity variable $u_k$.
There are also two types of auxiliary Bethe roots, or $y$-roots. Such roots transform in the fundamental representation of a different $su(2)$ algebra, sometimes called $su(2)_{\bullet}$~\cite{Borsato:2014exa}. Hence we introduce an index $\alpha=1,2$, and excitation numbers~$N_y^{(\alpha)}$. The (complex) rapidity variable for these particles is $y_k$; we will see later that a real rapidity~$u_k$ may also be introduced, depending on where in the complex plane $y_k$~lies.

\paragraph{Left equations.} We consider any ``left'' magnon (with $M=+1$) with mirror momentum $\tilde p_k$, $k=1,\dots,  N_1$. We have
\begin{equation}
\label{eq:BYEL}
    \begin{aligned}
    &1=
  e^{i\tilde{p}_kR}   \prod_{\substack{j=1\\j\neq k}}^{N_{{1}}}
  S_{sl}^{11}(u_k,u_j)
    \prod_{j=1}^{N_{\bar1}}
\widetilde{S}_{sl}^{11}(u_k,u_j)
\prod_{\dot{\alpha}=1}^2\prod_{j=1}^{N_0^{(\dot{\alpha})}} S^{10}(u_k,u_j^{(\dot{\alpha})})
    \prod_{\alpha=1}^2
    \prod_{j=1}^{N_y^{(\alpha)}} S^{1y}(u_k,y^{(\alpha)})\,,
    \end{aligned}
\end{equation}
where we have introduced the S~matrices
\begin{equation}
\label{eq:mixedmassSL}
\begin{aligned}
{S}_{sl}^{11}(u_k, u_{j}) &= \frac{x^+_k-x^-_j}{x^-_k-x^+_j}
    \frac{1-\frac{1}{x^-_kx^+_j}}{1-\frac{1}{x^+_kx^-_j}}\big(\sigma^{\bullet\bullet}_{kj}\big)^{-2}\,,
    \\
    {\St}_{sl}^{11}(u_k, u_{j}) &= e^{ip_k}
    \frac{1-\frac{1}{x^+_kx^+_j}}{1-\frac{1}{x^-_kx^-_j}}
    \frac{1-\frac{1}{x^-_kx^+_j}}{1-\frac{1}{x^+_kx^-_j}}\big(\widetilde{\sigma}^{\bullet\bullet}_{kj}\big)^{-2}\,,
    \\
   {S}^{10}(u_k, u_{j})   &= e^{-\frac{i}{2} p_k}e^{-ip_j}
    \frac{1-x^+_kx_j}{x^-_k-x_j}\big(\sigma^{\bullet\circ}_{kj}\big)^{-2}\,,
\end{aligned}
\end{equation}
which depend on the dressing factors introduced in~\cite{Frolov:2021fmj},%
\footnote{Here the dressing factors have been continued to the mirror kinematics. The explicit expression of these S~matrices is given in appendix~\ref{app:Smatrices} in terms of improved dressing factors, which allow us to write compact formulae valid for the ``fused'' dressing factors that we will introduce later.
}
as well as the auxiliary S~matrix
\begin{equation}
    {S}^{1y}(u_k, y_{j}) = e^{-\tfrac{i}{2}p_k}\frac{1-\frac{1}{x^-_ky_{j}}}{1-\frac{1}{x^+_ky_{j}}} =e^{\tfrac{i}{2}p_k}\frac{x^-_k-\frac{1}{y_{j}}}{x^+_k-\frac{1}{y_{j}}}\,.
\end{equation}

\paragraph{Right equations.} We consider now any ``right'' magnon with mirror momentum $\tilde{p}_k$, $k=1,\dots, N_{\bar{1}}$. We have
\begin{equation}
    \begin{aligned}
    &1=
 e^{i\tilde p_kR}  \prod_{\substack{j=1\\j\neq k}}^{N_{\bar1}}
    S_{su}^{11}(u_k,u_j)
    \prod_{j=1}^{N_1}
     \widetilde{S}_{su}^{11}(u_k,u_k)\prod_{\dot{\alpha}=1}^2
    \prod_{j=1}^{N_0^{(\dot{\alpha})}}
    \overline{S}^{10}(u_k,u_j^{(\dot{\alpha})})
    \prod_{\alpha=1}^2
   \overline{S}^{1y}(u_k,y^{(\alpha)})\,,
    \end{aligned}
\end{equation}
where the scattering between momentum-carrying modes is
\bal
{S}_{su}^{11}(u_k, u_{j}) &= e^{+i p_k}e^{-i p_j}
    \frac{x^-_k-x^+_j}{x^+_k-x^-_j}
    \frac{1-\frac{1}{x^-_kx^+_j}}{1-\frac{1}{x^+_kx^-_j}}\big(\sigma^{\bullet\bullet}_{kj}\big)^{-2}
    \\
  {\St}_{su}^{11}(u_k, u_{j}) &= e^{-i p_j}
    \frac{1-\frac{1}{x^-_kx^-_j}}{1-\frac{1}{x^+_kx^+_j}}
    \frac{1-\frac{1}{x^-_kx^+_j}}{1-\frac{1}{x^+_kx^-_j}}\big(\tilde{\sigma}^{\bullet\bullet}_{kj})^{-2}
    \\
    {\Sb}^{10}(u_k, u_{j}) &= e^{+\frac{i}{2} p_k}e^{-ip_j}
    \frac{x^-_k-x_j}{1-x^+_kx_j}\big(\sigma^{\bullet\circ}_{kj}\big)^{-2}
\eal
while the auxiliary S matrix is
\begin{equation}
    {\Sb}^{1y}(u_k, y_{j})= e^{-\tfrac{i}{2}p_k}\frac{x^+_k-y_j}{x^-_k-y_{j}}={1\ov {S}^{1y}(u_k, 1/y_{j})}\,.
\end{equation}

\paragraph{Massless equations.}
We now consider a massless mode, which could have $su(2)_\circ$ flavour $\dot{\alpha}=1,2$. Let us pick for definiteness $\dot{\alpha}=1$, so that the mirror momentum is any $\tilde{p}_k$, $k=1,\dots , N_0^{(1)}$. We have
\begin{equation}
    \begin{aligned}
    &-1=
  e^{i\tilde p_kR}\prod_{\substack{j=1\\j\neq k}}^{N_{0}^{(1)}}S^{00}(u_k^{(1)},u_j^{(1)})
  \prod_{j=1}^{N_{0}^{(2)}}
  S^{00}(u_k^{(1)},u_j^{(2)})
   \prod_{j=1}^{N_{{1}}}
   S^{01}(u_k,u_j)
     \prod_{j=1}^{N_{\bar1}}
     \overline{S}^{01}(u_k^{(1)},u_j)
   \\
    &\qquad\times  
    \prod_{\alpha=1}^2\prod_{j=1}^{N_y^{(\alpha)}}
    S^{0y}(u_k^{(1)},y^{(\alpha)})
   \,.
    \end{aligned}
\end{equation}
Note that, had we chosen a massless particle with $\dot{\alpha}=2$, we would have obtained the same equation up to trivially relabeling the product limits in the first line. The S matrices here introduced are
\bal
{S}^{00}( u_k, u_{j})  &=  \big(\sigma^{\circ\circ}_{kj}\big)^{-2}\,,
    \\
   {S}^{01}(u_k, u_{j})  &= e^{+ip_k}e^{+\frac{i}{2} p_j}
    \frac{x_k-x^-_j}{x_kx^+_j-1}\big(\sigma^{\circ\bullet}_{kj}\big)^{-2}={1\ov {S}^{10}( u_j, u_{k})}\,,\\
   {\Sb}^{01}(u_k, u_{j}) &= e^{+ip_k}e^{-\frac{i}{2} p_j}
    \frac{x_kx^+_j-1}{x_k-x^-_j}\big(\sigma^{\circ\bullet}_{kj}\big)^{-2}={1\ov {\Sb}^{10}( u_j, u_{k})}\,,
\eal
and
\begin{equation}
    {S}^{0y}(u_k, y_{j})  = e^{-\tfrac{i}{2}p_k}\frac{x_k-y_{j}}{\frac{1}{x_k}-y_{j}}={1\ov  {S}^{0y}(u_k, {1\ov y_{j}}) }\,.
\end{equation}

\paragraph{Auxiliary equations.}
We finally consider the equation for any auxiliary root $y_{k}^{(\alpha)}$ with $ k=1,\dots ,N_y^{(\alpha)} $ and  $\alpha =1,2$. It does not depend explicitly on~$\alpha$ and it takes the form
\begin{equation}
\label{eq:BYEaux}
   - 1=
    \prod_{j=1}^{N_{{1}}}
    S^{y1}(y_k,u_j)
    \prod_{j=1}^{N_{\bar1}}
    \bar{S}^{y1}(y_k,u_j)
    \prod_{\dot{\alpha}=1}^2\prod_{j=1}^{N_0^{(\dot{\alpha})}}
    S^{y0}(y_k,u_k^{(\dot{\alpha})})\,.
\end{equation}
These S matrices follow from braiding unitarity,
\begin{equation}
     {S}^{y1}( y, p)=  {1\ov {S}^{1y}(p,y)}\,,\qquad
     {\Sb}^{y1}( y, p)=  {1\ov {\Sb}^{1y}(p, y)}\,,\qquad
     {\Sb}^{y0}( y, p)=  {1\ov {\Sb}^{0y}(p, y)}\,.
\end{equation}
\subsection{Bound states and fusion}
\label{sec:boundstates}
We now want to  consider the complex configurations of momenta that may appear in the thermodynamic limit. We know that massive particles may create bound states in the mirror theory~\cite{Frolov:2021fmj}, similar to those appearing in the string theory~\cite{Borsato:2013hoa}. As always, while the total (mirror) momentum and energy of a bound state are real, such an object is actually composed of several constituents with complex momenta and energies, arranged in a string-like pattern in some appropriate parametrisation, symmetrically with respect to the real~line.

\paragraph{Bound states of massive particles.} In $AdS_3\times S^3\times T^4$ we must distinguish between left bound states, which sit in representations with $u(1)$ charge $M$ given by a positive integer $Q$, and right bound states that have $u(1)$ charge given by a negative integer $-\overline{Q}$~\cite{Borsato:2013hoa}. We can think that they are made up of $Q$ left fundamental particles (each contributing charge $+1$), and of $\overline{Q}$ right fundamental particles (each contributing charge $-1$), respectively.
The mass gap for those two types of bound states is therefore $Q$ and $\overline{Q}$, respectively. In what follows we will refer to this as $Q$-particle and $\overline{Q}$-particle bound states. The massive fundamental particles are therefore special cases where $Q=1$ or $\overline{Q}=1$.
It is easy to confirm this analysis by studying the singularities of the Bethe-Yang equations for complex mirror momenta and in the limit~$R\to\infty$. In our notation, the analysis of left bound states very closely resembles that of $AdS_5\times S^5$ as presented in \textit{e.g.}~\cite{Arutyunov:2009zu}: there are poles and zeros in the left-left S~matrix which compensates the zeros and poles coming from the term~$e^{i\tilde{p}_kR}$ for $R\to\infty$.
By way of example, let us briefly discuss a two-particle bound state. We can create such a left-left bound state exploiting the pole in~$S^{11}_{sl}(u_1,u_2)$ when $x_1^-=x_2^+$, see~\eqref{eq:mixedmassSL}. In the equation for~$\tilde{p}_1$~\eqref{eq:BYEL} this gives a solution as long as $e^{i\tilde{p}_1R}\to0$, \textit{i.e.}~$\Im[\tilde{p}_1]>0$, which is indeed possible. Similarly, in the equation for the second particle we would have~$e^{i\tilde{p}_2R}\to\infty$ and a zero in the S~matrix.
For right-right bound states, the analysis is slightly more involved: the right-right S~matrix has poles and zeros which, rather than compensating the behaviour of~$e^{i\tilde{p}_kR}$, makes it more singular. To fulfill the Bethe-Yang equations we need additional zeros or poles that come from the auxiliary equations. The reason for this complication is that the right-right equations are written in the $su(2)$ grading, rather than $sl(2)$ grading. We would have found a very similar expression in $AdS_5\times S^5$ had we insisted on writing the mirror Bethe-Yang equations in the $su(2)$ grading.
One may also wonder if we may create left-right bound states. However, looking at \textit{e.g.}~$S^{1\overline{1}}_{sl}(u_1,u_2)$ we see that the only zeros and poles appear when the imaginary parts of $x_1$ and $x_2$ have opposite signs, meaning that it is impossible to have both particles in the mirror region.

\paragraph{Lack of bound states involving    massless particles.}
Massless particles have a rather different kinematics with respect to massive particles.  The massless-massless scattering does not have poles or zeros, and it appears that physical massless excitations have only real momentum~\cite{Frolov:2021fmj}. However, the mixed-mass S~matrices have apparent poles, which might indicated the presence of bound states between mass-one and massless particles, see \textit{e.g.}~\eqref{eq:mixedmassSL}. Let us look more closely at that case. We see that we may have a pole in $S^{10}(u_1,u_2)$ if $x_1^-=x_2$, which by braiding unitarity gives a zero in $S^{01}(u_2,u_1)$. This could lead to a solution of the Bethe-Yang equations at large~$R$ as long as $\Im[\tilde{p}_1]>0$ and $\Im[\tilde{p}_2]<0$.
The corresponding mirror momentum is (see appendix~\ref{app:definitions})
\begin{equation}
    \tilde{p}=\frac{h}{2}\left(
    x^-_1 - \frac{1}{x^-_1}- x^+_1+\frac{1}{x_1^+}+\frac{2}{x_2}-2x_2
    \right)=
    \frac{h}{2}\left(\frac{1}{x_2}-x_2
     -x^+_1+\frac{1}{x_1^+}
    \right)\,.
\end{equation}
while the mirror energy is
\begin{equation}
    \widetilde{\mathcal{E}}=
    \log \frac{x^-_1}{x^+_1}\frac{1}{(x_2)^2}=\log\frac{1}{
    x^+_1\,x^-_1}=\log\frac{1}{
    x^+_1\,x_2}\,.
\end{equation}
This is formally the dispersion of a particle with $X^+=x^+_1$ and $X^-=1/x_2=1/x_1^-$. However, we must also require that the total energy is positive, which becomes a constraint on the rapidity $u_1$ of the original particle. We find that this is never possible, and hence we conclude that there is no such a bound state. A similar discussion holds for right-massless bound states.

\paragraph{Auxiliary roots.}
Finally, we have the auxiliary roots $y$. In general, and in particular in $AdS_5\times S^5$~\cite{Arutyunov:2009zu}, one could consider Bethe strings composed of several auxiliary roots, again in a string-like configuration on a suitable rapidity plane. By repeating the analysis of~\cite{Arutyunov:2009zu} for the case at hand, however, it is immediate to see that such configurations cannot exist here. Unlike $AdS_5\times S^5$ (or the closely-related Hubbard model), here we have only one type of auxiliary roots, the $y$-roots, which are not sufficient to create a Bethe string configuration. In fact, in order to satisfy the auxiliary Bethe-Yang equations~\eqref{eq:BYEaux}, any root $y_k$ should have $|y_k|=1$; otherwise, we would obtain that either all auxiliary S~matrices $S^{y1}(u_j,y_k)$, $S^{y\bar{1}}(u_j,y_k)$ and $S^{y0}(u_j,y_k)$ have modulus smaller than one (if $|y_k|>1$), or all have modulus larger than one (if $|y_k|<1$). Either way, we would not be able to solve the auxiliary Bethe-Yang equation~\eqref{eq:BYEaux}.
At a more fundamental level, this is due to the fact that the auxiliary roots only encode a $su(1|1)^{\oplus4}$  centrally-extended symmetry, which does not have any non-Abelian Bosonic sub-algebra --- as opposed to, for instance, $su(2|2)$ or $su(2|2)^{\oplus2}$ centrally extended. In summation, we do not need to consider strings of $y$-roots in the thermodynamic limit.

\bigskip
Let us now discuss in some more detail the structure of the $Q$- and $\bQ$-particles and of the auxiliary roots.

\paragraph{Left-left bound states, or $Q$-particles.}
Owing to our choice of the $sl(2)$ grading for the left equations, the description of these bound states is particularly simple. The constituents are precisely~$Q$ left momentum-carrying roots satisfying
\bal
x_1^-=x_2^+\,,\quad x_2^-=x_3^+\,,\quad\ldots\, \quad x_{ Q-1}^-=x_Q^+\,,
\eal
where the Zhukovsky variables can be written on the $u$-plane as
\bal
x_j^\pm = x\big(u_j\pm{i\ov h}\big) \,,\qquad u_j = u +\frac{(Q+1-2j)\,i}{h}\,,\qquad j=1,\dots Q\,,\quad u\in\bR\,.
\eal
It is easy to check that such a configuration solves the mirror Bethe-Yang equations.

\paragraph{Right-right bound states, or  $\bQ$-particles.}
Here, as we anticipated, the description is a little more involved because the right Bethe-Yang equations are in the $su(2)$ grading. Hence we have to consider $\bQ$ right momentum-carrying modes, as well as $2(\bQ-1)$ auxiliary roots. The momentum-carrying roots obey
\bal\la{Qpartcond}
x_1^-=x_2^+\,,\quad x_2^-=x_3^+\,,\ldots\, \quad x_{\bQ -1}^-=x_{\bQ}^+
\eal
with
\bal
x_j^\pm = x\big( u_j\pm{i\ov h}\big) \,,\qquad  u_j =  u +\frac{(\bQ+1-2j)i}{h}\,,\qquad
j=1,\dots\bQ\,,\quad
u\in\bR\,.
\eal
The auxiliary roots $y_{j}^{(1)}$ and $y_{ j}^{(2)}$ are
\bal\la{Qpyroos}
y_{j}^{(1)}=y_{j}^{(2)}={1\ov x_j^-} = {1\ov x( u +( \bQ-2j){i\ov h})}\,,\qquad j=1,2,\ldots,  \bQ-1\,,
\eal
and they  satisfy the complex-conjugation condition
\bal
{1\ov \big(y_{j}^{(\a)}\big)^*} =  \Big(x\big( u +( \bQ-2j){i\ov h}\big)\Big)^* = {1\ov  x( u -( \bQ-2j){i\ov h})} = {  y_{\bQ-j}^{(\a)}}\,.
\eal
It is not hard to see, following~\cite{Arutyunov:2009zu}, that such complexes solve the mirror Bethe-Yang equations --- essentially, the poles in the auxiliary S~matrices compensate the zeros in the right-right S~matrix.

\paragraph{Massless particles.}
Massless modes do not form bound states, and only appear as individual excitations of real (mirror) momentum and energy. This means that they have~\cite{Frolov:2021fmj}
\begin{equation}
    -1< x^{(\dot{\alpha})}<1\,,\qquad \dot{\alpha}=1,2\,.
\end{equation}

\paragraph{Auxiliary roots.}
Finally, we come to the $y$-roots, which as we stressed cannot create strings. Instead, we have individual roots, each lying on the unit circle
\begin{equation}
    \big(y^{(\alpha)}\big)^* = \frac{1}{y^{(\alpha)}}\,,\qquad\alpha=1,2\,,
\end{equation}
as it can be easily seen by the arguments of~\cite{Arutyunov:2009zu}.

\subsection{Fusion of the Bethe-Yang equations}
\label{sec:fused}
We now want to recast the Bethe-Yang equations by combining the constituents of $Q$- and $\bQ$-particles. The resulting equations will feature ``fused'' S~matrices, given by the product over their constituents. Typically, this product telescopes at least partially, yielding fairly manageable S~matrices, see \textit{e.g.}~\cite{Arutyunov:2009zu}. This is fairly straightforward for the rational pieces of the S~matrices, but rather subtle for the dressing factors. For the dressing factors of $AdS_3\times S^3\times T^4$ this was discussed in~\cite{Frolov:2021fmj}, while for the Beisert-Eden-Staudacher factor~\cite{Beisert:2006ez} (which plays an important role here too) it was discussed in~\cite{Arutyunov:2009kf}. Below we collect the various ``fused'' equations, while the ``fused'' S~matrices appearing here are defined in appendix~\ref{app:Smatrices}, and the fused dressing factors are defined in appendix~\ref{app:factors}.

\paragraph{Left equations ($Q$ particles).}
We write here the equation for any $Q_a$-particle, meaning for a  bound state of $Q_a$ left magnons. Each such bound state has real mirror momentum $\tilde{p}_a$, and its (fused) Zhukovsky variable is given by
\begin{equation}
\label{eq:boundZhukovsky}
    x_a^\pm = x\big(u_a\pm{i\ov h}Q_a\big)\,.
\end{equation}
In a generic state, we will have $N_L$ such bound states, so that $a=1,\dots, N_L$. The Bethe-Yang equation for the $a$-th bound state is
\bal \la{BEQfused} 
  1=
    e^{i\tilde p_aR}&\prod_{\substack{b=1\\ b\neq a}}^{N_{ L}}{S}_{sl}^{Q_aQ_b}(  u_a, u_{b}) 
    \prod_{b=1}^{N_{R}}{\St}_{sl}^{Q_a \bQ_b}(  u_a, u_{b})
    \prod_{\dot{\alpha}=1}^2\prod_{j=1}^{N_0^{(\dot{\alpha})}}{S}^{Q_a 0}(  u_a, x_{j}^{(\dot{\alpha})})
     \prod_{\alpha=1}^2\prod_{b=1}^{ N_y^{(\alpha)}}{S}^{Q_a y}( u_a, y_{b}^{(\alpha)})\,.
\eal
Here $N_L$ and $N_R$ are the number of left and right magnons and bound states, respectively, while $N_y^{(\a)}$ now stands for the number of single auxiliary roots which are not in any (right) bound state; finally, $N_0^{(\dot{\alpha})}$ are, like before, the numbers of massless momentum-carrying modes. The explicit form of these S~matrices can be found in appendix~\ref{app:Smatrices}.

\paragraph{Right equations ($\bQ$-particles).}
In a similar way we can derive the equation for a $\bQ_a$-particle, which is made up of $\bQ_a$ right particles and $2(\bQ_a-1)$ auxiliary excitations. In a generic state, we have $N_R$ such bound states, so that $a=1,\dots N_R$. Each of them has real mirror momentum and its Zhukovsky variables are given by~\eqref{eq:boundZhukovsky}. The Bethe-Yang equations are
\bal\la{BEbQfused}
 1=e^{i \tilde  p_a R}& 
\prod_{\substack{b=1\\ b\neq a}}^{N_R} S_{su}^{\bQ_a \bQ_b}( u_a, u_b) 
    \prod_{b=1}^{N_L}{\St}_{su}^{\bQ_aQ_b}(  u_a, u_{b})
    \prod_{\dot{\alpha}=1}^2\prod_{j=1}^{N_0^{(\dot{\alpha})}}{\Sb}^{\bQ_a0}(  u_a, x_{j}^{(\dot{\alpha})})   
    \prod_{\alpha=1}^2\prod_{b=1}^{N_y^{(\alpha)}}{\Sb}^{\bQ_ay}( u_a, y_{b}^{(\alpha)})\,.
\eal

\paragraph{Massless equations.}
Consider now the equation for a  massless particle. For definiteness, we take $\dot{\alpha}=1$. We have $N_{0}^{(1)}$ such particles with real mirror momentum $\tilde{p}_a$, $a=1,\dots N_{0}^{(1)}$ (as well as $N_0^{(2)}$ particles with $\dot{\alpha}=2$). They scatter with the various bound states we described above, giving
\bal\la{BE0fused}
  &-1=
    e^{i\tilde p_kR}\prod_{\substack{j=1\\j\neq k}}^{N_0^{(1)}}(\phase^{\circ\circ}_{kj}\big)^{-2}
    \prod_{j=1}^{N_0^{(2)}}(\phase^{\circ\circ}_{kj}\big)^{-2}
    \prod_{b=1}^{N_L}{S}^{0Q_b}(  x_k, u_b)
    \prod_{b=1}^{N_R}{\Sb}^{0\bQ_b}(  x_k, u_b)
    \prod_{\alpha=1}^2
    \prod_{b=1}^{N_y^{(\alpha)}}{S}^{0y}( x_k, y_{b}^{(\alpha)}) .
\eal

\paragraph{Bethe-Yang equation for auxiliary roots.}
Finally we have the equation for any auxiliary root $y_{k}^{(\alpha)}$ , where the flavour $\alpha=1,2$ and $ k=1,\dots N_y^{(\alpha)} $. It can be written in terms of the S matrices already introduced, and it reads
\begin{equation}\la{BEauxfused}
   - 1=
    \prod_{b=1}^{N_L}S^{yQ_b}(y_k^{(\alpha)},u_b)
    \prod_{b=1}^{N_{R}}\overline{S}^{y\overline{Q}_b}(y_k^{(\alpha)},u_b)
    \prod_{\dot{\alpha}=1}^2\prod_{j=1}^{N_0^{(\dot{\alpha})}}S^{y0}(y_k^{(\alpha)},x_j^{(\dot{\alpha})}) \,.
\end{equation}

\subsection{String hypothesis}
\label{sec:hypothesis}

The rewriting  of the Bethe-Yang equations of the  previous section is useful in the thermodynamic limit. This is when the mirror volume~$R$ is very large, $R\to\infty$, and the various excitation numbers $N$ are also very large, $N\to\infty$, while the densities  $N/R$ are finite.
In that case we have argued that, like in~\cite{Arutyunov:2009zu}, all relevant configurations of particles are captured by four types of composite objects, each having real mirror momentum and energy, and namely
\begin{enumerate}
\item  One set of $N_L^{(Q)}$ of $Q$-particles with real momenta $\tp_k$, and $\sum_{Q=1}^\infty N_L^{(Q)} = N_L$,
\item  One set of $N_R^{(\bQ)}$ of  $\bQ$-particles with real momenta $\tp_k$,  and $\sum_{\bQ=1}^\infty N_R^{(\bQ)} = N_R$,
\item Two sets of $N_0^{(\dot{\alpha})}$ massless particles, of flavour $\dot{\alpha}=1,2$, with real momenta $\tp_k$ and real rapidity $x_k$ satisfying $-1<x_k^{(\dot{\alpha})}<1$,
\item  Two sets of $N_y^{(\a)}$ ``auxiliary particles'' $y^{(\a)}$ with flavour $\alpha=1,2$, corresponding to a auxiliary roots $y^{(\a)}$ with $|y^{(\a)}|=1$.
\end{enumerate}
The mirror energy of the resulting state is given by
\begin{equation}
    \widetilde{\mathcal{E}}=
    \sum_{a=1}^{N_L}\widetilde{\mathcal{E}}_{Q_a}(p_a) + 
    \sum_{a=1}^{N_R}\widetilde{\mathcal{E}}_{\overline{Q}_a}(p_a) + 
    \sum_{\dot{\alpha}=1}^2\sum_{k=1}^{N_0^{(\dot{\alpha})}}\widetilde{\mathcal{E}}_{0}(p_k^{(\dot{\alpha})})\,,
\end{equation}
where we remark that the auxiliary $y^{(\a)}$ particles do not contribute, and the explicit form of the mirror dispersion is given in appendix~\ref{app:definitions}.

In conclusion, in the thermodynamic limit the scattering problem is reduced to the diagonal scattering of various types of ``particles'' of real momentum. The inverted commas here are due to the fact that the $y^{(\a)}$ excitations are not true particles, or at least they are not momentum-carrying modes and they do not contribute to the energy of the state. Regardless, we may characterise them by a real rapidity and quite straightforwardly derive the TBA equations for their densities, as we will do in the next section.

\section{Mirror TBA equations} 
\label{sec:mirrortba}
Once the string hypothesis is formulated, the derivation of the TBA equations follows a standard route. We need to introduce densities for the roots (as well as for holes) for each of the ``particles'' (in the generalised sense explained at the end of the previous section) appearing in the string hypothesis. Since each of these particles is real, their densities will live on the real line of a suitable rapidity plane. In our case, this will be the $u$ plane (see appendix~\ref{app:definitions}), with real particles living on (a subset of) the real-$u$ line. Unfortunately, and much like in $AdS_5\times S^5$, some of the scattering matrices have branch points in the $u$-plane, including on the real $u$-line for $u=\pm2$. Hence some particles will live on the whole real-$u$ line, while others will live on a certain segment --- either within the interval $(-2,2)$, or outside it.

Then, the TBA equations follow by rewriting the logarithm of the mirror Bethe-Yang equations in the thermodynamic limit (introducing integrals and convolutions) and demanding that the free energy is stationary. The resulting TBA equations will involve a number of kernels (which we will typically indicate with $K$), which follow from the S~matrices introduced above.

While the TBA equations are a complete and correct description of the spectrum in and of themselves, they are fairly unwieldy objects as they involve infinite sums and convolutions of infinitely many objects (the densities, which are typically repackaged in terms of so-called $Y$-functions). For this reason it is often convenient to simplify the TBA equations, as we do at the end of this section. We will see that, much like in $AdS_5\times S^5$ (see~\cite{
Bombardelli:2009ns,Gromov:2009bc,Arutyunov:2009ur, Arutyunov:2009ux,Cavaglia:2010nm}), the equations cannot be simplified quite as completely as we would expect for, say, a relativistic model.

\subsection{Integral equations for densities}
\label{sec:densities}

In the thermodynamic limit we introduce densities $\rho(u)$ of particles, and densities $\br(u)$ of holes which depend on the real rapidity variable~$u$ (the bar on the densities is unrelated to the $\overline{Q}$ particles introduced above). We have the following types of densities, some of which depend on the indices $\alpha=1,2$ and $\dot{\alpha}=1,2$.
\begin{enumerate}
\item The densities $\rho_Q(u)$ and $\rho_\bQ(u)$ of the $Q$- and $\bQ$-particles, respectively, 
\begin{equation}
    \rho_Q(u),\quad Q=1,2,\ldots \infty\,,\qquad 
    \rho_\bQ(u)\,,\quad\bQ=1,2,\ldots\infty\,,\qquad u\in\mathbb{R}\,.
\end{equation}
\item The density $\rho_{0}^{(\dot{\a})}(u)$ of the massless particles,
\begin{equation}
    \rho_{0}^{(\da)}(u)\,,\quad\dot{\a}=1,2\,,\qquad |u|>2\,.
\end{equation}
The corresponding $x$-coordinate is expressed in terms of $u$ as $x=x(u+ i0)$,
or as $1/x_s(u)$.
\item The density $\rho_{y^-}^{(\a)}(u)$ of the auxiliary $y$-particles with negative imaginary part
\begin{equation}
    \Im[y]<0:\qquad
    \rho_{y^-}^{(\a)}(u)\,,\qquad -2<u<2\,.
\end{equation}
The corresponding $y$-coordinate is expressed in terms of $u$ as $y=x(u)$, or as $y=x_s(u- i0)$.
\item The density $\rho_{y^+}^{(\a)}(u)$ of the $y$-particles with Im$(y)>0$, $-2\le u\le 2$. 
\begin{equation}
    \Im[y]>0:\qquad
    \rho_{y^+}^{(\a)}(u)\,,\qquad -2<u<2\,.
\end{equation}
The corresponding $y$-coordinate is expressed in terms of $u$ as $y=1/x(u)$, or as $y=x_s(u+ i0)$.
\end{enumerate}
For each of  these densities of roots, we introduce the corresponding density of holes, taking values in the same region of the $u$-plane.

Introducing a generalised index $i$ which runs over all the densities, we will be able to represent the system of integral equations arising in the thermodynamic limit in the following compact form
\begin{eqnarray}\la{tbacom}
\rho_i(u) + \br_i(u) = {R\ov 2\pi}{d\tp_i\ov du} + K_{ij}\star\rho_j (u)\,.
\end{eqnarray}
where the momentum $\tp_i$ does not vanish only for massive and massless particles.  Here and in what follows  the summation over $j$ is assumed, and the star product is defined by the following composition law
\bal\la{starp}
K_{ij}\star\rho_j (u) = \sum_j\int {\rm d}u'\, K_{ij}(u,u')\rho_j (u')\,,
\eal
where  the kernel $K_{ij}$ is defined through the corresponding S matrix by 
\bal
K_{ij}(u,v)={1\ov 2\pi i}{d\ov du}\log S_{ij}(u,v)\,,
\eal
and the integration is taken over the range of $u$ specified above for each of the various types of densities. When writing more explicit equations later on, we will use some special symbols for the convolution over the interval $(-2,2)$, or over the interval $(-\infty,-2)\cup(2,+\infty)$, namely
\begin{equation}
    \star \leftrightarrow\int\limits_{-\infty}^{+\infty}\text{d}u\,,\qquad
    \hat{\star} \leftrightarrow\int\limits_{-2}^{+2}\text{d}u\,,\qquad
    \check{\star} \leftrightarrow\Big(\int\limits_{-\infty}^{-2}+\int\limits_{+2}^{+\infty}\Big)\text{d}u\,.
\end{equation}
Regardless of the domain, the convolution \eqref{starp} can be thought of as the left
action of the kernel $K_{ij}$ on $\rho_j$.
In what follows we will
also need the right action which is defined as  
\bal\la{starp2}
\rho_j \star K_{ji}(u) = \sum_j\int {\rm d}u'\, \rho_j
(u')K_{ji}(u',u)\,,
\eal
and similarly for $\hat{\star}$ and $\check{\star}$.
We can now spell out the various equations more explicitly.

\paragraph{Equations for Q-particle densities.}
To derive the integral equation for $Q$-particle densities, we first
rewrite the Bethe-Yang equation \eqref{BEQfused} for
$Q$-particles  in the form
\bal\la{BEQ}
  1=&
    e^{i\tilde p_aR}\prod_{\substack{b=1\\ b\neq a}}^{N_{ L}}{S}_{sl}^{Q_aQ_b}(  u_a, u_{b})
    \prod_{b=1}^{N_{R}}{\St}_{sl}^{Q_a \bQ_b}(  u_a, u_{b})
    \prod_{\da=1,2}\,  \prod_{j=1}^{N_0^{(\da)}}{S}^{Q_a 0}(  u_a, u_{j}) 
    \\
    &\times \prod_{\a=1,2}\, \prod_{b=1}^{N_\a^-}{S}^{Q_a y}_-( u_a, u_{b}) 
    \prod_{b=1}^{N_\a^+}{S}^{Q_a y}_+( u_a, u_{b})\,,
\eal
where we have split the product over $y$-roots depending on their imaginary parts. 
Taking the logarithmic derivative of \eqref{BEQ} with respect to $u_a$, we get in the thermodynamic limit the following integral equation for the densities of $Q$-particles and holes
\begin{eqnarray}\la{tbaQ}
\rho_Q(u) + \br_Q(u) &=& {R\ov 2\pi}{d\tp^Q(u)\ov du} + \sum_{Q'=1}^\infty K_{sl}^{QQ'}\star\rho_{Q'}+ \sum_{\bQ'=1}^\infty \tK_{sl}^{Q\bQ'}\star\rho_{\bQ'} 
\\\nonumber
&&~~~~~~~~+  \sum_{\da=1,2} K^{Q0}\cstar\rho_{0}^{(\da)}+  \sum_{\a=1,2}\left[ K^{Qy}_-\hstar\rho_{y^-}^{(\a)}+ K^{Qy}_+\hstar\rho_{y^+}^{(\a)} \right] \,.~~~~
\end{eqnarray}
$|u|\ge2$ and $|u|\le2$, respectively. 

\paragraph{Equations for $\overline{\text{Q}}$-particle densities.}
Similarly, we
rewrite the Bethe-Yang equation \eqref{BEbQfused} for
$\bQ$-particles  in the form
\bal\la{BEbQ}
 1=e^{i \tilde  p_a R}& 
\prod_{\substack{b=1\\ b\neq a}}^{N_{R}} S_{su}^{\bQ_a \bQ_b}( u_a, u_b)   \prod_{b=1}^{N_{{L}}}{\St}_{su}^{\bQ_aQ_b}(  u_a, u_{b})
 \prod_{\da=1,2}\,  \prod_{j=1}^{N_0^{(\da)}}{\Sb}^{\bQ_a0}(  u_a, u_{j})   
    \\
    &\times \prod_{\a=1,2}\,  \prod_{b=1}^{N_\a^-}{1\ov S_+^{\bQ_ay}( u_a, u_{b})}  \prod_{b=1}^{N_3^+}{1\ov S_-^{\bQ_ay}( u_a, u_{b})}\,,
\eal
and get the  integral equation for the densities of $\bQ$-particles and holes
\begin{eqnarray}\la{tbQ}
\rho_\bQ(u) + \br_\bQ(u) &=& {R\ov 2\pi}{d\tp^\bQ(u)\ov du} + \sum_{\bQ'=1}^\infty K_{su}^{\bQ\bQ'}\star\rho_{\bQ'}+ \sum_{\bQ'=1}^\infty \tK_{su}^{\bQ Q'}\star\rho_{Q'} 
\\\nonumber
&&\qquad+ \sum_{\da=1,2} \tK^{\bQ0}\cstar\rho_{0}^{(\da)} -  \sum_{\a=1,2}\left[ K^{\bQ y}_+\hstar\rho_{y^-}^{(\a)}+ K^{\bQ y}_-\hstar\rho_{y^+}^{(\a)} \right] \,.
\end{eqnarray}

\paragraph*{Equations for massless particle densities.}
Next, we
rewrite the Bethe-Yang equation \eqref{BE0fused} for
massless particles  in the form
\bal
  1&=
    e^{i\tilde p_kR}\prod_{\substack{j=1\\j\neq k}}^{N_0^{(1)}}{S}^{00}(  u_k, u_{j})\prod_{j=1}^{N_0^{(2)}}{S}^{00}(  u_k, u_{j}) 
    \prod_{b=1}^{N_{{L}}}{S}^{0Q_b}(  u_k, u_{b})
    \prod_{b=1}^{N_{R}}{\Sb}^{0\bQ_b}(  u_k, u_{b})
    \\
    &\times \prod_{\a=1,2}\,  \prod_{b=1}^{N_\a^-}{1\ov {S}^{0y}( u_k, u_{b}) }
    \prod_{b=1}^{N_\a^+}{ {S}^{0y}( u_k, u_{b}) }\,,
\eal
where for definiteness we considered the case in which $\tilde{p}_k$ has flovour $\dot{\alpha}=1$. For either flavour $\da =1,2$ we get the  integral equation for the densities of massless particles and holes
\begin{eqnarray}\la{tba0}
\rho_0^{(\da)}(u) + \br_0^{(\da)}(u) &=& {R\ov 2\pi}{d\tp^0(u)\ov du} +K^{00}\cstar\rho_{0}^{(1)}+K^{00}\cstar\rho_{0}^{(2)} + \sum_{Q=1}^\infty K^{0Q}\star\rho_{Q}
\\\nonumber
&&\qquad + \sum_{\bQ=1}^\infty \tK^{0\bQ}\star\rho_{\bQ} + \sum_{\a=1,2}\left[ K^{0 y}\hstar(\, \rho_{y^+}^{(\a)} -\rho_{y^-}^{(\a)}\, )\right] \,.
\end{eqnarray}

\paragraph{Equations for y-particle  densities with $\text{Im(y)<0}$.}
Next, we consider a $y^{(\a)}$-particle with the root
$y^{(\a)}_k=x(u_k^{(\a)})$ and rewrite the equation \eqref{BEauxfused} in the following form
\bal\la{BEym}
-1=\prod_{b=1}^{N_{L}} S^{yQ}_-(u_k^{(\a)},u_b)\prod_{b=1}^{N_{R}}S^{y\bQ}_+(u_k^{(\a)},u_b)\prod_{j=1}^{N_0} S^{y0}(u_k^{(\a)},u_j)
\eal
and we get in the thermodynamic limit the following integral equation for the densities of $y^-$-particles and holes
\begin{eqnarray}\la{tbaym}
\rho_{y^-}^{(\a)}(u) + \br_{y^-}^{(\a)}(u) = \sum_{Q=1}^\infty K^{yQ}_-\star\rho_{Q} +\sum_{\bQ=1}^\infty K^{y\bQ}_+\star\rho_{\bQ} +\sum_{\da=1,2}K^{y0}\cstar\rho_{0}^{(\da)}   \,.
\end{eqnarray}

\paragraph{Equations for y-particle densities with $\text{Im(y)>0}$.}
Finally, in the second case ${\rm Im}(y)>0$ the root $y^{(\a)}_k=1/x(u_k^{(\a)})$ we rewrite the equation \eqref{BEauxfused} in the following form
\bal\la{BEyp}
-1=\prod_{b=1}^{N_{L}} S^{yQ}_+(u_k^{(\a)},u_b)\prod_{b=1}^{N_{R}}S^{y\bQ}_-(u_k^{(\a)},u_b)\prod_{j=1}^{N_0} S^{y0}(u_k^{(\a)},u_j)\,.
\eal
Taking the logarithmic derivative of (\ref{BEym}) with respect to $u_k^{(\a)}$, we get in the thermodynamic limit  the following integral equation for the densities of $y^+$-particles and holes
\begin{eqnarray}\la{tbayp}
\rho_{y^+}^{(\a)}(u) + \br_{y^+}^{(\a)}(u) = \sum_{Q=1}^\infty K^{yQ}_+\star\rho_{Q} +\sum_{\bQ=1}^\infty K^{y\bQ}_-\star\rho_{\bQ} +\sum_{\da=1,2}K^{y0}\cstar\rho_{0}^{(\da)}   \,.~~~~
\end{eqnarray}

\subsection{Ground-state TBA equations}
\label{sec:gstba}

Using the thermodynamic limit introduced above, the derivation of the ground state TBA equations is standard. It is customary and convenient to repackage the densities in terms of ``$Y$-functions'', which in our case take the form
\bal
Y_Q =  {\rho_Q\ov\br_Q}\,,\qquad  \bY_Q =  {\rho_\bQ\ov\br_\bQ}\,,\qquad Y_0^{(\da)} =  {\rho_0^{(\da)}\ov\br_0^{(\da)}}\,,\qquad Y_\pm^{(\a)}
= -  e^{i \mu_\a}\, {\bar\rho_{y^\pm}^{(\a)}\ov\rho_{y^\pm}^{(\a)}}\,,
\eal
where $\mu_\a=(-1)^\a\mu$, and $\mu$ is a twist parameter which if $\mu\neq0$ breaks supersymmetry of the light-cone string theory. 

In terms of these Y-functions, after imposing that the free energy is stationary (see for instance~\cite{vanTongeren:2016hhc} for a pedagogical review of the procedure), we get the following equations:
\paragraph{Q-particles.}
\bal\la{TbaQ}
-\log  Y_Q &= L\, \tE_{Q} - \log\left(1+Y_{Q'} \right)\star K_{sl}^{Q'Q} -  \log\left(1+ \bY_{Q'} \right)\star \tK_{su}^{Q' Q} 
\\
&- \sum_{\da=1,2} \log\left(1+Y_0^{(\da)} \right)\cstar K^{0Q} \\
 &- \sum_{\a=1,2}\log\left(1-{e^{i \mu_\a}\ov Y_{+}^{(\a)}} \right)\hstar K^{yQ}_+ - \sum_{\a=1,2} \log\left(1-{e^{i \mu_\a}\ov Y_{-}^{(\a)}} \right)\hstar K^{yQ}_-\,.~~~~~~
\eal

\paragraph{$\overline{\text{Q}}$-particles.}
\bal\la{TbabQ}
-\log   \bY_Q &= L\, \tE_{Q} - \log\left(1+ \bY_{Q'} \right)\star K_{su (2)}^{Q'Q} -  \log\left(1+Y_{Q'} \right)\star \tK_{sl}^{Q' Q} 
\\
&- \sum_{\da=1,2}   \log\left(1+Y_0^{(\da)} \right)\cstar \tK^{0Q} \\
 &- \sum_{\a=1,2}\log\left(1-{e^{i \mu_\a}\ov Y_{+}^{(\a)}} \right)\hstar K^{yQ}_- - \sum_{\a=1,2} \log\left(1-{e^{i \mu_\a}\ov Y_{-}^{(\a)}} \right)\hstar K^{yQ}_+\,.~~~~~~
\eal

\paragraph{Massless particles.}
\bal\la{Tba0}
-\log Y_0^{(\da)} &= L\, \tE_{0} - \sum_{\dot\beta=1,2}  \log\left(1+Y_{0}^{(\dot\beta)} \right)\cstar K^{00} -  \log\left(1+Y_{Q} \right)\star K^{Q 0} -  \log\left(1+ \bY_{Q} \right)\star \tK^{Q 0} \\
 &- \sum_{\a=1,2}\log\left(1-{e^{i \mu_\a}\ov Y_{+}^{(\a)}} \right)\hstar K^{y0} - \sum_{\a=1,2} \log\left(1-{e^{i \mu_\a}\ov Y_{-}^{(\a)}} \right)\hstar K^{y0}\,.~~~~~~
\eal

\paragraph{y$^{\boldsymbol -}$-particles.}
\bal
&\log Y_-^{(\a)} =  - \log\left(1+Y_{Q} \right)\star K^{Qy}_-  + \log\left(1+ \bY_{Q} \right)\star K^{Q y}_+   + \sum_{\da=1,2} \left( \log(1+Y_{0}^{(\da)} \right)\cstar K^{0y} \,.~~~~~~
\eal

\paragraph{y$^{\boldsymbol +}$-particles.}
\bal
&\log Y_+^{(\a)} =  - \log\left(1+Y_{Q} \right)\star K^{Qy}_+  + \log\left(1+ \bY_{Q} \right)\star K^{Q y}_-   -  \sum_{\da=1,2} \log\left(1+Y_{0}^{(\da)} \right)\cstar K^{0y} \,.~~~~~~
\eal

\paragraph{Ground-state energy.}
\bal
\la{energyL}
E(L) &=-\int\limits_{-\infty}^\infty {{\rm d}u\ov
2\pi}{d\tp^Q\ov du} \log\Big(\big(1+Y_Q \big)\big(1+ \bY_Q \big) \Big)\\
&\qquad\qquad\qquad\qquad-\int\limits_{|u|>2} {{\rm d}u\ov
2\pi}{d\tp^0\ov du}\log\Big(\big(1+Y_0^{(1)}  \big)\big(1+Y_0^{(2)}  \big)\Big)\,.
\eal
For $\mu=0$ the TBA equations describe the even-winding number sector of the light-cone string theory with periodic fermions and supersymmetric vacuum while for $\mu=\pi$ they describe the odd-winding number sector with anti-periodic fermions and nonsupersymmetric vacuum. For general values of $\mu$ fermions satisfy twisted boundary conditions.

Since for $\mu=0$ the TBA equations are written for the sector of states with a BPS vacuum, the ground state energy must vanish. This means that the BPS ground state $Y$-functions are given by 
\bal
Y_Q=\bY_Q=Y_0^{(\da)}=0\,,\quad  Y_\pm^{(\a)}=1\,,\quad \mu=0\,.
\eal
Strictly speaking, for $\mu=0$ the TBA equations for $Y_Q, \bY_Q$ and $Y_0^{(\da)}$ functions are singular, and a proper way to analyse them is to consider a small $\mu$ expansion. Such an analysis for the \ads case was performed in \cite{Frolov:2009in}, and 
it would be interesting to do the same for the $AdS_3\times S^3$ case. 

Let us also mention that just as it was in the \ads case, the functions $Y_-^{(\a)}$ and $Y_+^{(\a)}$ can be considered as two branches of one function $Y^{(\a)}$ defined on the Riemann surface obtained by gluing together the mirror and anti-mirror $u$-planes.

\subsection{Simplified TBA equations}
\label{sec:gstbasim}

We want now to simplify the TBA equations derived above and eliminate, to the extent that it is possible, the infinite sums over particle types $Q$ and $\bQ$ occurring in the right-hand side. For the equations of $Y_Q$ and $\bY_Q$ functions, this simplification follows a standard route. 
We start by introducing the kernel
\bal\la{invK0} \left( K  + 1\right)_{MN}^{-1} = \delta_{MN}
- s\left( \delta_{M+1,N}+ \delta_{M-1,N}\right)\,,\quad s(u)=
{g\ov 4\cosh {g\pi u\ov 2}}\,.
\eal 
that is inverse to the
kernel $ K_{NQ}  + \delta_{NQ}$ 
\bal \la{invKK}
\sum_{N=1}^\infty \left( K_{QN}  +
\delta_{QN}\right)\star\left( K + 1\right)_{NM}^{-1}= \delta_{QM}= \sum_{N=1}^\infty\left( K + 1\right)_{MN}^{-1} \star\left( K_{NQ}  +
\delta_{NQ}\right)\,.
\eal
Using various identities that this satisfies (see for instance~\cite{Arutyunov:2009ur,Arutyunov:2009ux}) we find the following equations
\bal
-\log Y_Q &=  \log\left(1+{1\ov Y_{Q-1}}\right)\left(1+{1\ov Y_{Q+1}}\right)\star s\,,\quad Q\ge 2
\\
-\log  \bY_Q &= \log\left(1+{1\ov   \bY_{Q-1}}\right)\left(1+{1\ov  \bY_{Q+1}}\right)\star s\,.
\eal
Very nicely, the dependence on $\overline{Y}_Q$ functions drops out of the equations for $Y_Q$ functions, and vice-versa. Coming to the equations for fundamental particles, we see that they couple to auxiliary roots. However, similarly to the case of $AdS_5\times S^5$, the equations for fundamental particles cannot be quite  put in the same form. We can rewrite them as
\bal
-\log Y_1 &=  \log\left(1+{1\ov Y_{2}}\right)\star s- \log\left(1-{e^{-i \mu}\ov Y_{-}^{(1)}} \right)\left(1-{e^{i \mu}\ov Y_{-}^{(2)}} \right)\hstar s  + F_1\cstar s\,,
\\
-\log  \bY_1 &= \log\left(1+{1\ov  \bY_{2}}\right)\star s - \log\left(e^{-i \mu}-{Y_{+}^{(1)}} \right)\left(e^{i \mu}-{ Y_{+}^{(2)}} \right)\hstar s  + \overline F_1\cstar s
\eal
where $F_1$ and $\overline F_1$ whose explicit forms are not important for the current discussion depend on all $Y$-functions and various kernels.

Using these expressions, we can gain insight on the form of the Y-system for this model. To this end, we introduce the operator $s^{-1}$ that acts on functions of the
rapidity variable $u$ as follows 
\bal\la{defs} (f\star s^{-1})(u)
= \lim_{\epsilon\to 0^+} \big[ f(u+{i\ov g} - i\epsilon )+ f(u-{i\ov g} + i\epsilon
)\big]\,.
\eal
Under convolution, $s^{-1}$ is a right inverse of $s$, but it is not a left inverse of $s$, meaning that at least for some function $f(u)$, $f\star s^{-1}\star s\neq f$. In other words, by applying $s^{-1}$ we loose some information from the TBA equations (which encodes the discontinuity relations between $Y$ functions). Applying $s^{-1}$ to the simplified equations we find Y-system equations 
\bal
{Y_Q^+Y_Q^-\ov Y_{Q-1}Y_{Q+1}}&={1\ov (1+Y_{Q-1})(1+Y_{Q+1})}\,,\quad Q\ge 2
\\
{\bY_Q^+\bY_Q^-\ov \bY_{Q-1}\bY_{Q+1}}&={1\ov (1+\bY_{Q-1})(1+\bY_{Q+1})},
\eal
valid on the real line, and 
\bal
{Y_1^+Y_1^-\ov Y_{2}}&={\left(1-{e^{-i \mu}\ov Y_{-}^{(1)}} \right)\left(1-{e^{i \mu}\ov Y_{-}^{(2)}} \right)\ov 1+Y_{2} }\,,
\\
{\bY_1^+\bY_1^-\ov \bY_{2}}&={\left(e^{-i \mu}-{ Y_{+}^{(1)}} \right)\left(e^{i \mu}-{ Y_{+}^{(2)}} \right)\ov 1+\bY_{2} }\,,
\eal
valid for $-2<u<2$. 
It is clear that there are no Y-system equations of the standard form for the remaining Y-functions. Still, we expect non-trivial relations upon analytic continuation of $Y$-functions.

\section{Conclusions and outlook}
\label{sec:conclusions}
In this paper we have derived for the first time the mirror thermodynamic Bethe ansatz equations for $AdS_3\times S^3\times T^4$. Unlike previous attempts to the same end~\cite{Bombardelli:2018jkj,Fontanella:2019ury,Cavaglia:2021eqr,Ekhammar:2021pys}, our treatment does not rely on any truncation or limit.
While some of the features of AdS3/CFT2 integrability have been long known~\cite{Borsato:2012ud,Borsato:2013qpa,Borsato:2014hja}, a key ingredient of our derivation has been the proposal of a new set of dressing factors~\cite{Frolov:2021fmj} with nice self-consistent properties in the mirror-mirror kinematics and under fusion --- something that was not obvious in previous proposals~\cite{Borsato:2013hoa,Borsato:2016xns}.

There are several obvious directions which should be now pursued. As the mirror TBA equations presented here are complete and non-perturbative in the string tension, they may be immediately  used to study the spectrum of excited states. This can be done quite straightforwardly by employing the contour-deformation trick~\cite{Dorey:1996re,Arutyunov:2009ax}, and immediately overcomes all issues related to massless excitations in the Bethe-Yang equations~\cite{Abbott:2015pps}.
It would firstly provide a prediction to check against perturbative computations. The best-understood setup to compute these would be at $h\gg1$, on the string-theory side. This would require a careful study of the apparent UV and IR divergences of the worldsheet model. While there are several results at one- and two-loops in the literature~\cite{Rughoonauth:2012qd,Sundin:2013ypa,Engelund:2013fja,Roiban:2014cia,Bianchi:2014rfa,Sundin:2016gqe}, some of these results are manifestly incompatible with integrability. A possible reason for this mismatch is that those computations involved both an IR and UV regulator resulting in an order-of-limits problem, see~\cite{Frolov:2021bwp} for a discussion of this point and of the mismatches.

It would also be very interesting to study the spectrum at $h\ll1$. While the massive dynamics will probably simplify in this limit, we expect the massless modes to give quite a non-trivial contribution, as it can be seen already from their dressing factors.
A detailed computation would offer new quantitative insights in the dual theory for this background. It would be particularly interesting to revisit the non-local Lagrangian proposed in~\cite{OhlssonSax:2014jtq} and compare its predictions with the mirror~TBA.

On the other hand, it would also be interesting to further simplify these equations and recast them in terms of a T-system, and eventually in terms of a quantum spectral curve. The final result of this simplification procedure may be related to the quantum spectral curve bootstrapped in~\cite{Cavaglia:2021eqr,Ekhammar:2021pys}, in which case it should clarify how the massless modes feature in that formalism. The advantage in this reformulation is that the quantum spectral curve should be a more efficient computational framework, and it may also help unveil some of the underlying symmetry structures of the theory.

A different question is whether it may be possible to extend the mirror TBA to more general backgrounds. In particular, as discussed in the introduction, mixed-flux backgrounds are integrable. The main stumbling block before undertaking this analysis is to construct the dressing factors for the mixed-flux background, which in turn requires a detailed understanding of its analytic properties. Encouragingly, it seems that the approach of~\cite{Frolov:2021fmj} may also be applicable to mixed-flux backgrounds~\cite{upcoming:mixed}, which would lay the basis for solving the spectral problem across the parameter space of $AdS_3\times S^3\times T^4$.

Finally, it would be interesting to study more general AdS3/CFT2 setups. A particularly interesting one is given by the $AdS_3\times S^3\times S^3\times S^1$ background, which has yet one more parameter --- the ratio~$\alpha$ of the curvature radius of the two spheres. This directly affects the symmetry algebra of the theory, which is given by $d(2,1;\alpha)\oplus d(2,1;\alpha)$. Despite some progress in the study of the S~matrix and Bethe-Yang equations for this model~\cite{Borsato:2012ud,Borsato:2012ss,Borsato:2015mma} the dressing factors remain elusive, though the approach of~\cite{Frolov:2021fmj} may apply there too. We hope to return to these questions in the near future.

\section*{Acknowledgements}
AS gratefully acknowledges support from the IBM Einstein Fellowship.

\appendix

\section{Kinematics and parametrisations}
\label{app:definitions}
The relation between the Zhukovsky variable $x$ and the rapidity $u$ is
\begin{equation}
     x(u) = \frac{1}{2} \left(u-i \sqrt{4-u^2}\right)\,,
\end{equation}
which has cuts on the real-$u$ line for $u>+2$ and $u<-2$ (``long'' cuts).
In terms of this variable, we write for massive particles
\begin{equation}
    x_j^\pm = x\big(u_j\pm{i\ov h}\big)\,,
\end{equation}
with real particles having real rapidity $u_j$; for massive particles we also have bound states, whose constituents have complex energy, momentum and rapidity. In particular, for a bound state of $Q$ particles, we will have $Q$ constituents with rapidities arranged along a string parallel to the imaginary axis,
\begin{equation}
x_j^\pm = x\big( u_j\pm{i\ov h}\big) \,,\quad  u_j =  u +( \bQ+1-2j){i\ov h}\,,\qquad
j=1,\dots Q\,,\quad u\in\bR\,.
\end{equation}
As discussed in the main text, by using fusion bound states can also be thought of as particles with a mass gap $Q\in\mathbb{N}$, and may be described collectively in terms of a rapidity $u_a$ by setting
\begin{equation}
x_a^\pm = x\big(u_a\pm \frac{i}{h} Q_a\big)\,.
\end{equation}

For massless particles we have
\begin{equation}
    x_j = x(u_j+i0)=\frac{1}{x(u_j-i0)}\,,
\end{equation}
where all physical particles are real (there are no massless bound states) and have rapidity just above the long cut. With this caveat, we may formally think of a massless particle as a $Q\to0$ limit of a bound state (though we stress that this is not a fusion identity as it is for $Q\geq1$).

It is also convenient to introduce a $u$-parametrisation for the auxiliary particles. Physical auxiliary particles (those appearing in the string hypothesis) have $|y|=1$. We distinguish the case where the imaginary part of $y$ is smaller or larger than zero. We have
\begin{equation}
    \Im[y]>0\,:\quad x(u)=y\,,\qquad
    \Im[y]<0\,:\quad \frac{1}{x(u)}=y\,,
\end{equation}
with $-2<u<+2$.

The mirror momentum for a $Q$-particle is given by
\begin{equation}
\begin{aligned}
\tilde p^Q(u) &
=\frac{h}{2}\left(
x(u - \tfrac{i}{h}Q)-\frac{1}{x(u - \tfrac{i}{h}Q)} -x(u + \tfrac{i}{h}Q)+\frac{1}{x(u + \tfrac{i}{h}Q)}
\right)
\\
&= h\, x(u - \tfrac{i}{h}Q) - h\, x(u + \tfrac{i}{h}Q) + i\,Q\,,
\end{aligned}
\end{equation}

while the mirror energy for a $Q$-particle is given by
\bal
 \tE^Q(u) = \log {x(u - \tfrac{i}{h}Q) \ov x(u + \tfrac{i}{h}Q) }\,.
\eal
These formulae are formally valid for massless modes too by setting~$Q\to0$. The auxiliary roots do not contribute to the the mirror momentum or energy.

Finally, when dealing with the explicit form of the dressing factors it is also useful to introduce $\gamma$-rapidity variables~\cite{Frolov:2021fmj}.
We define the massive $\g^\pm$ for a $Q$-particle so that they are complex conjugate to each other for real mirror momentum or equivalently for real $u$
\bal
\g^+(u)&=\log \left(+\frac{i \left(-1+x(u+{i\ov h}Q)\right)}{1+x(u+{i\ov h}Q)}\right)-\frac{ i \pi
   }{2} =\frac{1}{2} \log \left(-\frac{u+\frac{i}{h}Q-2}{u+\frac{i}{h}Q+2}\right)-\frac{i \pi }{2}
   \\
   \g^-(u)&=\log \left(+\frac{i \left(-1+x(u-{i\ov h}Q)\right)}{1+x(u-{i\ov h}Q)}\right)+\frac{ i \pi
   }{2} =\frac{1}{2} \log \left(-\frac{u-\frac{i}{h}Q-2}{u-\frac{i}{h}Q+2}\right)+\frac{i \pi }{2}
\eal
Note that this definition differs from the one mainly used in~\cite{Frolov:2021fmj} by a shift by $i\pi/2$, and it is particularly advantageous when working in the mirror kinematics.
Then, we define the massless $\g$ so that it is real for real mirror momentum and for values of $u$ on the upper edge of the long cut
\bal
   \g(u)&=\frac{1}{2} \log \left(-\frac{u-2}{u +2}\right)+\frac{i \pi }{2}\,.
\eal
Obviously,
\bal
\g^-(u)= \g(u-\frac{i}{h}Q)\,,\quad \g^+(u)= \g(u+\frac{i}{h}Q)-i\pi\,.
\eal
Notice also that 
\bal
 \g(u+i0)- \g(u-i0)=-i\pi\quad\text{for}\quad |u|>2\,.
\eal

\section{S~matrices and kernels}
\label{app:Smatrices}
Let us start by introducing a standard S matrix for bound states, namely
\bal
\la{sqq} 
S^{QQ'}(u-u')&= \frac{u-u' -{i\ov h}(Q+Q')}{u-u' +{i\ov h}(Q+Q')}  \frac{u-u' -{i\ov h}(Q'-Q)}{u-u' +{i\ov h}(Q'-Q)}\\
 &\qquad\qquad\times\prod_{j=1}^{Q-1}\left(\frac{u-u' -{i\ov h}(Q'-Q+2j)}{u-u' +{i\ov h}(Q'-Q+2j)}\right)^2  \, ,
 \eal
which is of difference form in terms of $u$. In particular
 \bal
\la{s1q} 
S^{1Q'}(u-u')&= \frac{u-u' -{i\ov h}(1+Q')}{u-u' +{i\ov h}(1+Q')}  \frac{u-u' -{i\ov h}(Q'-1)}{u-u' +{i\ov h}(Q'-1)} \, 
 \eal
Using the above expression, as well as the dressing factors collected in appendix~\ref{app:factors} below, we can define the following S matrices which appear in the fused Bethe Yang-equations.

\paragraph{Left-anything scattering.}
These are the scattering matrices where the first particle is of type ``left'' (\textit{i.e.}, a $Q$-particle):
\bal
S_{sl}^{Q_a Q_b}( u_a, u_b)&= S^{Q_aQ_b}(u_a-u_b)^{-1}\big(\Sigma^{Q_aQ_b}_{ab}\big)^{-2}\,, 
\eal
\bal
{\St}_{sl}^{Q_a \bQ_b}(  u_a, u_{b}) &=e^{ip_a} \frac{1-\frac{1}{x^+_ax^+_b}}{1-\frac{1}{x^-_ax^-_b}}
    \frac{1-\frac{1}{x^+_ax^-_b}}{1-\frac{1}{x^-_ax^+_b}}  \big(\tSi^{Q_a\bQ_b}_{ab}\big)^{-2}\,,
\eal
\bal
{S}^{Q_a0}(  u_a, x_{j})  =\,i\,e^{-\frac{i}{2} p_a}  \frac{x^+_ax_j-1}{x^-_a-x_j}
   {\big(\Sigma_\bes^{Q_a0}(x_a^\pm,x_j) \big)^{-2} \ov \PhiSG(\gamma_{aj}^{+\circ})\,\PhiSG(\gamma_{aj}^{-\circ})}\,,
   \eal
   \bal
     {S}^{Q_a y}( u_a, y_{b}) &=e^{\tfrac{i}{2}p_a}\,\frac{x_a^-- {y_{b}}}{x_a^+- {y_{b}}}\,.
\eal
\paragraph{Right-anything scattering.}
These are the scattering matrices where the first particle is of type ``right'' (\textit{i.e.}, a $\overline{Q}$-particle):
\bal
S_{su}^{\bQ_a \bQ_b}( u_a, u_b)
&=e^{ip_a}e^{-ip_b}\left(  {x_a^+-x_{b}^-\ov  x_a^--x_b^+ }\right)^{-2} S^{\bQ_a\bQ_b}(u_a-u_{b})^{-1}\left( \Si^{\bQ_a\bQ_b}_{ab}\right)^{-2} 
\eal
\bal
{\St}_{su}^{\bQ_aQ_b}(  u_a, u_{b})    &=e^{-ip_b}
  \frac{1-\frac{1}{x^-_ax^-_b}}{1-\frac{1}{x^+_ax^+_b}}
    \frac{1-\frac{1}{x^+_ax^-_b}}{1-\frac{1}{x^-_ax^+_b}} \big(\tSi^{\bQ_aQ_b}(u_a,u_b)\big)^{-2}
\eal     
 \bal
 {\bar S}^{\bQ_a0}(  u_a, x_{j})  &=i\,e^{+\frac{i}{2} p_a}  \frac{x^-_a-x_j}{x^+_ax_j-1}{\big(\Sigma_\bes^{\bQ_a0}(x_a^\pm,x_j) \big)^{-2} \ov \PhiSG(\gamma_{aj}^{+\circ})\,\PhiSG(\gamma_{aj}^{-\circ})}
\eal
\bal
 {\Sb}^{Q_ay}( u_a, y_{b})&= e^{-\tfrac{i}{2}p_a}\frac{x^+_a-{1\ov y_b}}{x^-_a-{1\ov y_{b}}}={1\ov {S}^{Qy}( u_a, 1/y_b)}
 \eal
\paragraph{Massless-anything scattering.}
These are the scattering matrices where the first particle is massless:
\bal
\label{eq:S00}
S^{00}(u_j,u_k)=a(\gamma_{jk})\,\varPhi(\gamma_{jk})^2\ \big(\Sigma_\bes^{00}(x_j,x_k) \big)^{-2}
\eal
\bal
{S}^{0Q_b}(  x_b, u_{j})  {S}^{Q_b0}( u_j, x_b)=1\,,\quad {\Sb}^{0\bQ_b}(  x_b, u_{j})  {\Sb}^{\bQ_b0}( u_j, x_b)=1
\eal
\bal
 {S}^{0y}( x_k, y_{j})  &= e^{+\tfrac{i}{2}p_k}\frac{{1\ov x_k}-{ y_{j}}}{{x_k}-{ y_{j}}}={1\ov  {S}^{0y}( x_k, {1\ov y_{j}}) }
 \eal
\paragraph{Auxiliary-anything scattering.}
The S matrices where the first particle is auxiliary can be obtained from the previous ones by braiding unitarity,
\begin{equation}
     {S}^{yQ}( y, u)=  {1\ov {S}^{Qy}(u,y)}\,,\qquad
     {\Sb}^{yQ}( y, u)=  {1\ov {\Sb}^{Qy}(u, y)}\,,\qquad
     {\Sb}^{y0}( y, u)=  {1\ov {\Sb}^{0y}(u, y)}\,.
\end{equation}

\paragraph{Kernel.}
We can obtain the integration kernel for any S~matrix by first parametrising it in terms of the rapidity~$u$ and considering the logarithmic derivative,
\begin{equation}
    K(u,u')=\frac{1}{2\pi i}\frac{\text{d}}{\text{d}u}\log S(u,u')\,.
\end{equation}
In this way we will be able to define the various kernels, whose notation will follow from the one of the corresponding S matrix. For instance, we will define $K^{QQ'}_{sl}(u,u')$, $\widetilde{K}^{Q\overline{Q}'}_{sl}(u,u')$, $K^{\overline{QQ}'}_{su}(u,u')$, and so on.

\section{Dressing factors}
\label{app:factors}
The dressing factors have been recently proposed in~\cite{Frolov:2021fmj}. All of the dressing factors feature the BES dressing factor~\cite{Beisert:2006ez} in a suitable kinematics.
In~\cite{Frolov:2021fmj}, we focused our attention mostly on the string-theory dressing factors~$\sigma_{\text{BES}}^{QQ'}$, while in this paper it is more convenient to work with the ``improved'' dressing factors~$\Sigma_{\text{BES}}^{QQ'}$, which have nice properties under fusion in the mirror model. They are related by
\begin{equation}
\begin{aligned}
 \Sigma_\bes^{QQ'}(u,u') &=
\sigma_\bes^{QQ'}(u,u')\prod_{k=1}^Q\prod_{j=1}^{Q'} {1- {1\ov x_k^+x_{j}^-}\ov {1-{1\ov x_k^-x_{j}^+} } }\\
& =
\sigma_\bes^{QQ'}(u,u'){1- {1\ov y_1^+y_{2}^-}\ov {1-{1\ov y_1^-y_{2}^+} } }\prod_{k=1}^{Q-1} {1- {1\ov x_k^-y_{2}^-}\ov {1-{1\ov x_k^-y_{2}^+} } } \prod_{j=1}^{Q'-1} {1- {1\ov y_1^+x_{j}^-}\ov {1-{1\ov y_1^-x_{j}^-} } }\,,
\end{aligned}
\end{equation}
when $Q\geq1$ and ${Q'}\geq1$~\cite{Arutyunov:2009kf}.
For $Q=0$ we have~\cite{Frolov:2021zyc}
\begin{equation}
\Sigma_\bes^{0Q'}(u,u') =
\sigma_\bes^{0Q'}(u,u')\prod_{j=1}^{Q'}
\frac{\frac{1}{x}-x^-_j}{x-x^-_j}\,,
\end{equation}
while for massless-massless scattering we have simply
\begin{equation}
\Sigma_\bes^{00}(u,u') =
\sigma_\bes^{00}(u,u')\,.
\end{equation}

The improved mirror-mirror dressing factor $\Sigma^{QQ'}_{\text{BES}}$ was given in~\cite{Arutyunov:2009kf},
\begin{equation}
\begin{aligned}
&{1\ov i}\log\Sigma^{QQ'}_\bes (y_1,y_2)
 =
\Phi(y_1^+,y_2^+)-\Phi(y_1^+,y_2^-)-\Phi(y_1^-,y_2^+)+\Phi(y_1^-,y_2^-)\\
&\qquad\qquad-{1\ov 2}\left(\Psi(y_1^+,y_2^+)+\Psi(y_1^-,y_2^+)-\Psi(y_1^+,y_2^-)-\Psi(y_1^-,y_2^-)\right) \\
&\qquad\qquad+{1\ov
2}\left(\Psi(y_{2}^+,y_1^+)+\Psi(y_{2}^-,y_1^+)-\Psi(y_{2}^+,y_1^-)
-\Psi(y_{2}^-,y_1^-) \right)
 \\
&\qquad\qquad+{1\ov i}\log\frac{ i^{Q}\,\Gamma\big[Q'-{i\ov
2}h\big(y_1^++\frac{1}{y_1^+}-y_2^+-\frac{1}{y_2^+}\big)\big]} {
i^{Q'}\Gamma\big[Q+{i\ov
2}h\big(y_1^++\frac{1}{y_1^+}-y_2^+-\frac{1}{y_2^+}\big)\big]}{1-
{1\ov y_1^+y_2^-}\ov 1-{1\ov
y_1^-y_2^+}}\sqrt{\frac{y_1^+y_2^-}{y_1^-y_2^+}} \,.
\end{aligned}
\end{equation}
This expression can also be used to find the~$\Sigma^{0Q'}_\bes$ and~$\Sigma^{00}_\bes$ phases, as discussed in~\cite{Frolov:2021fmj}.

The remaining dressing factors (in the mirror region) are defined using $\Sigma^{QQ'}_{\text{BES}}$ and are
\bal
\Sigma^{QQ'}_{12}\,,\qquad Q\,,\, Q'=1,2,\ldots\,,\qquad  \widetilde{\Sigma}^{QQ'}_{12}\,,\qquad Q\,,\, Q'=1,2,\ldots\,.
\eal
The massive dressing factors, with $Q,Q'=1,2,...$  are given by  
\begin{equation}
\begin{aligned}
 \big(\Sigma^{QQ'}_{12}\big)^{-2} =   &
 -\frac{\sinh\tfrac{\gamma^{-+}_{12}}{2}}{\sinh\tfrac{\gamma^{+-}_{12}}{2}}
 e^{\varphi^{\bullet\bullet}(\gamma^\pm_1,\gamma^\pm_2)}\ \big(\Sigma_\bes^{QQ'}(x_1^\pm,x_2^\pm)\big)^{-2}\,,
    \\
    \big(\widetilde{\Sigma}^{QQ'}_{12}\big)^{-2}=&
    +\frac{\cosh\tfrac{\gamma^{+-}_{12}}{2}}{\cosh\tfrac{\gamma^{-+}_{12}}{2}}
    e^{\tilde{\varphi}^{\bullet\bullet}(\gamma^\pm_1,\gamma^\pm_2)}\ \big(\Sigma_\bes^{QQ'}(x_1^\pm,x_2^\pm)\big)^{-2}\,.
\end{aligned}
\end{equation}
The functions~$\varphi^{\bullet\bullet}$ and $\tilde{\varphi}^{\bullet\bullet}$ are given in~\cite{Frolov:2021fmj}. Because these prefactors are given by the difference of the rapidities~$\gamma_1^\pm$ and~$\gamma_2^\pm$, they take the same form in the mirror and string theory, much like it happens in relativistic models.

Let us now consider the massive-massless dressing factors. Since they do not fuse well, it is more convenient to work directly with the full S-matrix elements $S^{Q0}(u,u')$ and $S^{0Q}(u',u)$ which are spelled out in appendix~\ref{app:Smatrices}. It is worth emphasising that their form is different from the one they have in the string region see~\cite{Frolov:2021fmj}. This is because, in the convention for the $\gamma^\pm$ parameters which we are using and which guarantees $(\gamma^\pm)^*=\gamma^\mp$ (see appendix~\ref{app:definitions}) they are related to the string ones as $\gamma_{(s)}^\pm=\gamma^\pm-i\pi/2$, while for the massless ones it is $\gamma_{(s)}=\gamma+i\pi/2$. As a result, starting from the string-kinematics formula~\cite{Frolov:2021fmj}
 \bal
 {\Sb}^{\bQ0}(  u_{(s)}, x_{(s)}) &=i\sqrt{\frac{x^+_{(s)}}{x^-_{(s)}}}  \frac{x^-_{(s)}-x_{(s)}}{1-x^+_{(s)}x_{(s)}}{\coth\tfrac{\gamma_{(s)}^{+\circ}}{2}}{\coth\tfrac{\gamma_{(s)}^{-\circ}}{2}}\\
 &\qquad\times\PhiSG(\gamma_{(s)}^{+\circ})\,\PhiSG(\gamma_{(s)}^{-\circ})
 \Big(\Sigma_\bes^{\bQ 0}(x_{(s)}^\pm,x_{(s)})
    \Big)^{-2},
\eal
and using the crossing equation
\begin{equation}
    \varPhi(\g)  \varPhi(\g)=\,i\,\coth\frac{\g}{2}\,,
\end{equation}
we find that the factors $\varPhi(\gamma)$ invert and that the rational prefactor is suitably modified, so that
 \bal
 {\bar S}^{\bQ_a0}(u_a, x_{j})  =i\,\sqrt{\frac{x^+}{x^-}}
 \frac{x^--x}{x^+x-1}\frac{\Big(\Sigma_\bes^{\bQ_a0}(x^\pm,x) \Big)^{-2}}{ \PhiSG(\gamma^{+\circ})\,\PhiSG(\gamma^{-\circ})}\,,
\eal
as in appendix~\ref{app:Smatrices}.

Finally, the massless-massless scattering factor~\eqref{eq:S00} can be expressed straightforwardly in terms of~$\Sigma^{00}_{\bes}$, of the Sine-Gordon factor~$\varPhi(\gamma)$, and of an auxiliary function~$a(\gamma)$, see~\cite{Frolov:2021fmj}. Since in this case the latter two depend on the difference of two rapidities with the same transformation rules from the string to the mirror region, they appear in the same way in the two kinematic regions. 

\bibliographystyle{JHEP}
\bibliography{refs}
\end{document}